\documentclass[conference]{IEEEtran}
\usepackage{cite}
\usepackage{amsmath,amssymb,amsfonts}
\usepackage{algorithmic}
\usepackage{graphicx}
\usepackage{textcomp}
\usepackage{xcolor}
\usepackage{fancyhdr}
\usepackage[hyphens]{url}
\usepackage{times}
\usepackage{tikz}
\usepackage{amsmath}
\usepackage{xspace}
\usepackage{amsfonts}
\usepackage{microtype}
\usepackage{graphicx}
\usepackage{subfigure}
\usepackage{booktabs} % for professional tables
\usepackage{graphicx}
\usepackage{tikz}
\usepackage{amsmath}
\usepackage[english]{babel}
\usepackage{blindtext}
\usepackage{bbm}
\usepackage{textcomp}
\usepackage{ulem}
\usepackage{tabularx}
\usepackage{enumitem}
\usepackage{xcolor}
\usepackage{hyperref}
\usepackage[font=footnotesize,labelfont=bf]{caption}
\usepackage{amsmath}

\def\BibTeX{{\rm B\kern-.05em{\sc i\kern-.025em b}\kern-.08em
    T\kern-.1667em\lower.7ex\hbox{E}\kern-.125emX}}

\newcommand{\iscasubmissionnumber}{NaN}
\def\compactify{\itemsep=2pt \topsep=2pt \partopsep=1pt \parsep=1pt \leftmargin=1.6em}
\let\latexusecounter=\usecounter

\definecolor{azureC}{RGB}{32,117,184}
\definecolor{ec2C}{RGB}{246,133,54}

\newcommand{\azure}{\textcolor{azureC}{Azure}\xspace}
\newcommand{\ectwo}{\textcolor{ec2C}{EC2}\xspace}

\newcommand{\cmpilogo}{%
  \begingroup\normalfont
  \includegraphics[height=\fontcharht\font`\B]{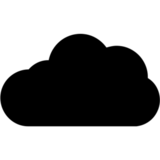}%
  \endgroup
}

\newcommand{\mpiC}{Collectives\xspace}
\newcommand{\mpi}{collectives\xspace}
\newcommand{\collectives}{\mpi}

\newcommand{\cmpi}{\cmpilogo\xspace{}\mpiC}
\fancypagestyle{firstpage}{
  \fancyhf{}

  \fancyhead[C]{\normalsize{ISCA 2021 Submission
      \textbf{\#\iscasubmissionnumber} \\ Confidential Draft: DO NOT DISTRIBUTE}} 
  \fancyfoot[C]{\thepage}
}

\title{\cmpi: Towards Cloud-aware \mpiC for ML Workloads with Rank Reordering}

\author{Liang Luo$^{*\ddagger}$,  Jacob Nelson$^\dagger$, Arvind Krishnamurthy$^*$, Luis Ceze$^*$ \\
$^*$University of Washington, $^\dagger$Microsoft Research, $^\ddagger$Facebook AI}
\begin{document}
\maketitle
\thispagestyle{firstpage}
\pagestyle{plain}

%%%%%% -- PAPER CONTENT STARTS-- %%%%%%%%

\begin{abstract}
%-------------------------------------------------------------------------------
ML workloads are becoming increasingly popular in the cloud. Good cloud training performance is contingent on efficient parameter exchange among VMs. We find that \mpiC, the widely used distributed communication algorithms, cannot perform optimally out of the box due to the hierarchical topology of datacenter networks and multi-tenancy nature of the cloud environment.

In this paper, we present \cmpi (Cloud-aware \mpi), a prototype that accelerates \mpi by reordering the ranks of participating VMs such that the communication pattern dictated by the selected \mpi operation best exploits the locality in the network. \cmpi is non-intrusive, requires no code changes nor rebuild of an existing application, and runs without support from cloud providers. Our preliminary application of \cmpi on \textit{allreduce} operations in public clouds results in a speedup of up to 3.7x in multiple microbenchmarks and 1.3x in real-world workloads of distributed training of deep neural networks and gradient boosted decision trees using state-of-the-art frameworks.

\end{abstract}

\section{Introduction}
Collective communication operations are an essential component of many data-parallel computation frameworks. 
Originally developed for high-performance computing frameworks such as MPI ~\cite{walker1996mpi}, they are now widely used for cloud-based distributed machine learning~\cite{vishnu2016distributed, WritingD41:online, ExtendMX17:online,NIPS2016_6381,Ke2017LightGBMAH, SparkMPI16:online, hpcincloud} workloads.
With increasingly more complex models~\cite{AIandCom71:online, TuringNL73:online, rajbh2019zero} calling for larger data exchanges, and rapid deployment of faster accelerators~\cite{sagemaker,brainwave,Jouppi:2017:IPA:3079856.3080246} demanding more frequent exchanges, efficient execution of these workloads contingent on efficient \mpi.

Unfortunately, achieving good \mpi performance in a cloud environment is fundamentally more challenging than in an HPC world, because the user has no control over node placement, topology and has to share the infrastructure with other tenants. These constraints have a strong implication on the performance of \mpi. As a result, the bottleneck of running these workloads on the cloud has shifted from computation to communication~\cite{7092922,PLink,phub}.

\begin{figure}[ht]
    \centering
    \includegraphics[width=.7\linewidth, trim=8 3 14 10,clip]{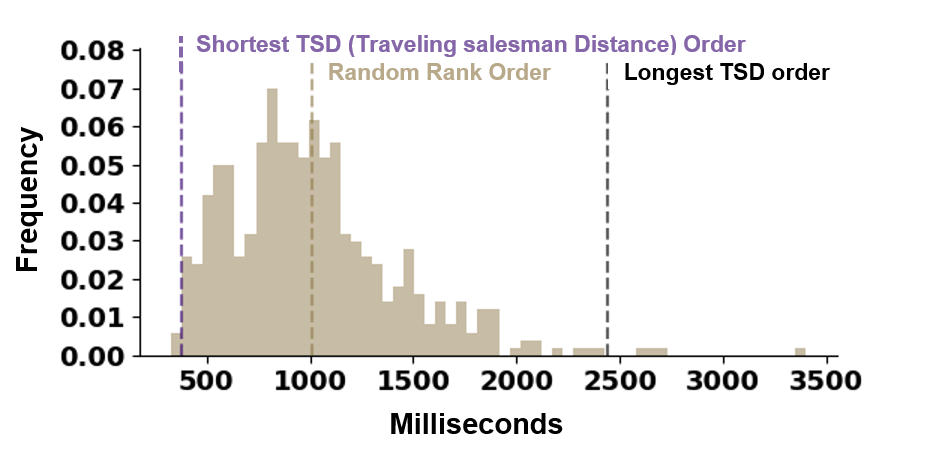}
    \caption{Performance distribution of \textit{allreduce} task of 100MB data with ring algorithm varies widely with 500 random rank orders on \azure.}
    \label{fig:azringperformance}
\end{figure}

\begin{figure}[ht]
    \centering
    \includegraphics[width=.6\linewidth]{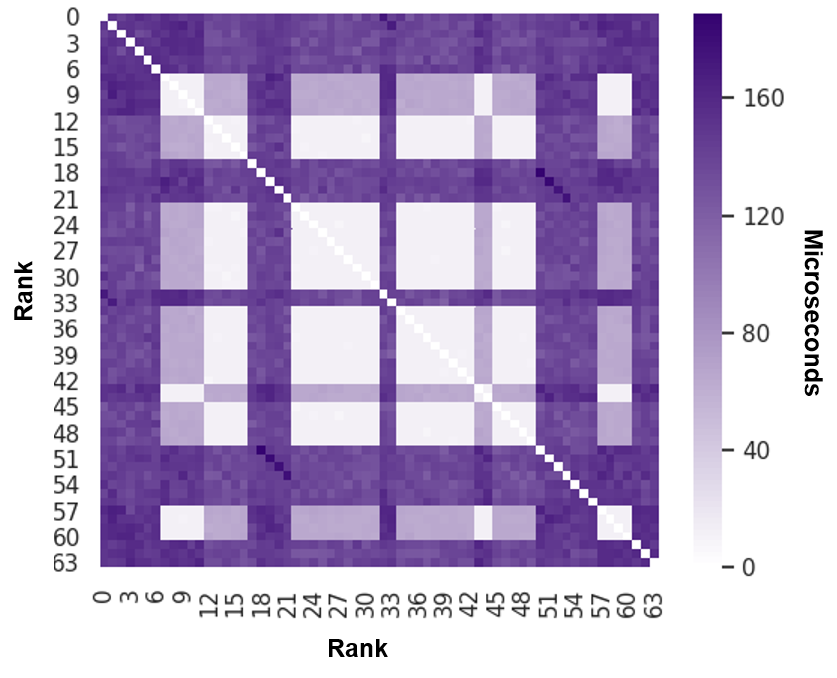}
    \caption{Pairwise RTT probe on 64 \azure Standard F64 v2 VMs shows non-uniform latency in point to point VM communication.}
    \label{fig:azpairwiseprobe}
\end{figure}

Consider a common practice of applying \textit{allreduce} ring \mpi, a popular algorithm, in the cloud context, where a randomly-ordered IP list (obtained through the provider) of VMs is used to form a virtual ring on which data is passed along, with $i$-th VM sending data to $i+1$-th VM. But do different ways of forming ring (through permutation of VMs in the list) exhibit the same performance? The answer is most likely no, as the ring that corresponds to shorter total hop cost will likely perform better~(Figure~\ref{fig:azringperformance}). On the other hand, not all ways of forming rings achieve the same cost, because the \textit{point to point communication cost (bandwidth, latency, or collectively referred to as locality in this work) is different across VMs}~(Figure~\ref{fig:azpairwiseprobe}), due to the hierarchical structure of the datacenter network, and the dynamic nature of traffic from other tenants. Consequently, running \mpi with a randomly-ordered list of VMs results in unpredictable and subpar performance. 

Our work focuses on discovering a permutation of the IP list that exploits the network locality for efficient communication, in a completely transparent way, by minimizing the cost model of a given \mpi parameterized with the actual hop cost. To do so, we need to (1) efficiently identify the underlying network constraints (or collectively, locality); (2) accurately build cost model for the \mpi at hand; (3) effectively approximate the minimum of complex cost functions.

This paper proposes \cmpi, a tool that uses network probes to discover locality within the underlying datacenter network, and uses it to solve a communication cost minimization problem with constraints, with the rank of each VM as the unknowns. We use reordered ranks as input to unmodified communication backends in microbenchmarks including OMB ~\cite{10.1007/978-3-642-33518-1_16}, Nvidia NCCL~\cite{NVIDIACo76:online}, Facebook Gloo~\cite{goyal2017accurate} and
real-world workloads of training deep neural networks with Pytorch/Caffe2 and gradient boosted decision trees using LightGBM~\cite{NIPS2016_6381,Ke2017LightGBMAH} and our preliminary results show a speedup of up to 3.7x in various \textit{allreduce} operations and 1.3x in end-to-end performance across \ectwo and \azure.

\section{Background}
\label{sec:background}
We provide an overview of typical structures and performance implications for datacenter networks and a brief introduction to the various popular \mpi operations.

\subsection{Locality in Datacenter Network}
Modern datacenters are hierarchical, with machines connecting to a top of rack switch, which are in turn connected to upper-level devices~\cite{Mysore2009PortLandAS,VL2,Roy2015InsideTS,incbricks}. This particular topology induces locality~\cite{PLink}, as the communication performance between two physical hosts is not the same. For example, VMs within the same rack have the best and stable performance, as the physical link is not shared. On the other hand, links between hosts residing in different racks are shared, and the communication performance depends on factors like hop count, link congestion, oversubscription ratio~\cite{Bilal2012ACS}, and dynamic load. Topology information is crucial for achieving optimal performance as many \mpi implementations generate routines based on this information~\cite{4154092,1639364,4228133,1419910,phub}. But in a cloud-environment, this information is hidden. Various attempts are made to reconstruct the physical affinity, e.g., PLink~\cite{PLink} uses DPDK-based latency probes and K-Means clustering to find hosts with high physical affinity.

%More importantly, we will show a full understanding of actual physical topology is not necessary for a cloud environment, because with multiple tenants sharing the same hardware in the cloud, better static physical affinity does not always equal higher performance, as dynamic traffic load ultimately decides the cost of each link.

\subsection{\mpiC}

\mpiC works by decomposing an operation into a series of point to point operations between two nodes according to a predefined communication schedule. \mpi most often appear in MPI contexts~\cite{Sack:2011:SCM:2522220,mpich,collectivesOptimization,blum2000architectures,bala1995ccl}. Typically, all nodes in \mpi participate in the communication, usually running symmetric tasks. \mpi can be used in many tasks such as \textit{(all)reduce, broadcast, (all)gather, and (reduce)scatter}~\cite{pagestac0:online}, and it is thus impractical to individually optimize each task. Fortunately, many of the tasks can be decomposed into multiple stages of \collectives primitives (e.g., \textit{allreduce} can be decomposed into \textit{reducescatter} followed by \textit{allgather}. Therefore, we only need to focus on accelerating such primitives. We now introduce these algorithms. We use $N$ as the number of participating nodes, $S$ as the amount of data to process per node.

\begin{figure}[t!]
    \centering
    \includegraphics[width=.9\linewidth]{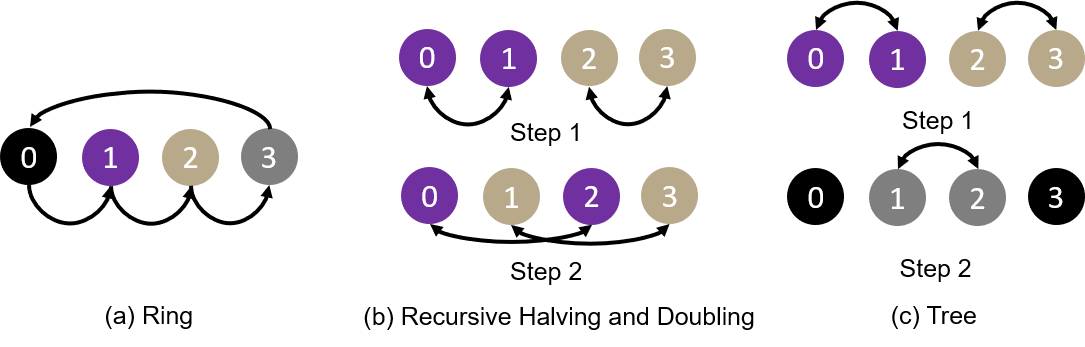}
    \caption{Rounds of communications (color-coded) in various popular algorithms used in \mpi with 4 nodes.}
    \label{fig:collectivesAlgorithm}
\end{figure}

\noindent\textbf{Ring}~\cite{patarasuk2009bandwidth}. As shown in Figure~\ref{fig:collectivesAlgorithm}(a), ring algorithms work by connecting nodes to form a virtual ring. Data is then passed along the ring sequentially. Ring algorithms require $O(N)$ steps to complete, sending $O(NS)$ amount of data.

\noindent\textbf{Halving Doubling}~\cite{mpich}. As shown in Figure~\ref{fig:collectivesAlgorithm}(b), halving doubling works by recursively doubling the distance (in terms of rank ID) while halving the total amount of data sent in each round, requiring $O(log_{2}{N})$ steps to finish while sending $O(NS)$ amount of data on the wire.

\noindent\textbf{Tree}. In one form, a single tree is built where data is transferred from leaves to the root and vice-versa~\cite{firecaffe}; in a more optimized setting, a pair of complementary binary trees are built to fully utilize the full bisection bandwidth~\cite{dbt}, each sending and receiving $S/2$. For binary trees, $O(log_2N)$ rounds of communication are required, sending $O(NS)$ bytes. 

\noindent\textbf{BCube}~\cite{glooalgo70:online}. BCube is very similar to halving doubling from a structural perspective, in the sense that nodes are organized into a group of $B$ peers. BCube operates in $O(log_BN)$ rounds, and each node in each round would peer with a unique node in another $B-1$ groups. Each node communicates $\frac{S}{B^i}$ amount of data in round $i$. BCube achieving a total bytes on wire of $O(\sum_{i=0}^{log_BN-1}\frac{S}{B^{i+1}})$.
\section{Motivation}

This section motivates \cmpi by demonstrating the implication of rank order on the performance of cloud-based \mpi.
We start by highlighting the asymmetric, non-uniform link cost in the cloud environment, by launching 64 Standard F64 VMs on the \azure cloud. We then run an in-house hybrid DPDK and ping-based~\cite{HomeDPDK76:online} latency probe (\S \ref{sec:probinglatency}) between each pair of VM node, using the same technique in PLink. The result is summarized in Figure~\ref{fig:azpairwiseprobe} as a heatmap. We observe the pairwise round-trip latency can range from sub-10 to hundreds of microseconds. 

We proceed to examine the performance of \textit{allreduce} operation using Gloo~\cite{GitHubfa54:online}, running the Ring chunked algorithm with 512 Standard F16 VMs. To derive a performance distribution, we use 500 randomly generated rank orders to generate 500 samples, and each is the average runtime of a 10-iteration reduction of 100MB of data. The result is summarized in the yellow distribution in Figure~\ref{fig:azringperformance}. The performance of different rank orderings of VM varies drastically, ranging from 330ms to 3400ms, with a mean of 1012ms and a standard deviation of 418ms. Now we have established the profound influence of rank ordering of VM nodes on the performance of \collectives algorithms, the goal of this paper is to derive an approximately optimal rank-ordering given a selected \collectives algorithm such that it maximizes the performance.

\section{Design and Implementation}
\label{sec:designandimpl}
We now describe \cmpi, a tool that takes in a list of VM nodes and a target algorithm, accurately and efficiently probes their pairwise distance, and uses that information to construct a rank order of VMs that attempts to minimize the total cost of communication.

\subsection{Cost Models for Collective Algorithms}
\cmpi builds a cost model $\mathbb{C}_\mathbb{O}$ for each popular algorithm used in \mpi $\mathbb{O}$, parameterized with the number of participating nodes $N$ and size $S$. This section details the cost models for popular algorithms. We use $c_{i,j}(S)$ to refer to the cost for transferring $S$ amount of data from node $i$ to $j$. We further define $MAX_{i=0}^{j}(f(i)) = MAX(f(0),...,f(j))$. We assume $N$ a power of 2 to simplify explanation, and allow arbitrary rank $r$ to alias to canonical rank $(r + N) \text{ mod } N$.
 
\noindent\textbf{Ring}. The cost model of the ring algorithm is the sum of the cost of each hop when traversing the ring: 

$$\mathbb{C}_{r}(N, c, S) = \sum_{i=0}^{N-1} c_{i,i-1}(S)$$

\noindent\textbf{Having Doubling}. The cost of halving doubling is the sum of costs for each round of communication, which in turn is the max cost of all communications in that round. 

$$\mathbb{C}_{hd}(N, c, S) = \sum_{i=0}^{log_2N-1}MAX_{j=0}^{\frac{N}{2}-1}c_{j,j+2^i}(\frac{S}{2^{i+1}}) $$

\noindent\textbf{Tree}. The cost of running tree algorithms depends on the number of trees and how trees are constructed. The total cost is the maximum cost of all trees, which is in turn determined by the maximum cost of each subtree. We provide a cost model for a popular variant of tree algorithm: double binary tree as used in~\cite{Massivel64:online}.
$$\mathbb{C}_{dbt}(N, c, S) = T(0,N-1,S)$$
where $T(i,j,S)$ is expressed recursively: 
\begin{align*}
 T(i,j,S) = 
 \left\lbrace
\begin{array}{l@{}l}
0 \textbf{ if $i \ge j$}\\
MAX(c_{\frac{i+j}{2}, \frac{3i+j}{2}-1}(\frac{S}{2}) + T(i, \frac{i+j}{2}-1), \\
c_{\frac{i+j}{2}, \frac{i+3j}{2}+1}(\frac{S}{2}) + T(\frac{i+j}{2}+1, j)) \textbf{ otherwise}
\end{array}
\right.
\end{align*}
Similarly a mirrored tree is built by decrementing each node's rank in the tree without changing the tree structure.

\noindent\textbf{BCube}. The cost of running the BCube algorithm is similar to halving doubling, except in each round, each node communicates with $B-1$ peers, instead of 1.

{\small
$$\mathbb{C}_{b}(N, c, S, B) = \sum_{i=0}^{log_BN-1}MAX_{j=0}^{\frac{N}{B}-1}MAX_{k=1}^{B}c_{j,j+kB^i}(\frac{S}{B^{i+1}}) $$
}

\subsection{Probing for Pairwise Distance}
\label{sec:probinglatency}
We need to determine values for $c_{i,j}(S)$ with end-to-end measurements. In this work, we use a latency-centric view for the cost component. The rationale behind this stems from the well-known theoretical TCP bandwidth model of $BW=O(\frac{MSS}{RTT\sqrt{p}})$ ~\cite{mathis1997macroscopic}: given constant drop rate $p$ and window $MSS$, higher latency induces lower bandwidth in TCP streams. This conveniently lets us approximate costs by only probing for latency. We adopt the probing pipeline used in PLink, which focuses on discovering physical locality with an in-house DPDK based echo tool, leveraging network enhancement provided by the clouds~\cite{Createan37:online, Enablean80:online}. Each pair of nodes receive a total of $10k$ probes from sequentially and bidirectionally. To derive an accurate reading, we take the RTT of 10th percentile to filter out interference during probes. Each probe is a UDP packet with a 32-bit payload that encodes sequence number and round id for fault tolerance. When DPDK cannot be used, we use \textit{fping}, a ICMP Echo-based latency probing tool. For each entry in $c$, we update $c_{i,j} \leftarrow MAX(c_{i,j}, c_{j,i})$ to make it symmetric.

\subsection{Minimizing the Cost Model}
We parameterize the cost model with values of probed $c$. To derive a rank ordering that minimizes $\mathbb{C_O}$, we perform the following transformation: let set of variables $\mathbb{R}$ defined as $r_i, i \in [0,N-1]$ be a permutation of $[0,N-1]$ to be solved, and we replace each $c_{i,j}$ with $c_{r_i,r_j}$. We can then establish a bijection from the original rank ordering to the desired order $r_i \leftrightarrow i$ once $r_i$s are solved. We flatten $c_{i,j} \leftrightarrow c'_{iN + j}$ to use theory of arrays to allow direct solving with conventional optimizing SMT solvers such as Z3~\cite{de2008z3,ORToolsG24:online}. 

Unfortunately, we find solvers inefficient, perhaps due to the non-convex, non-linear nature of the objective function and a large search space ($N!$). Thus, we take a two-stage solving process. The first step employs a range of stochastic search techniques such as simulated annealing~\cite{simanneal}, with a few standard heuristics (e.g., permuting a random sub-array, permuting random pairs) for obtaining neighboring states and a timeout. When the search returns with an initial result $C_0$, we generate an additional SMT constraint $\mathbb{C_O} < C_0$ to better guide pruning for solvers. We let the solver continue to run for a few minutes, and we either find a better solution or will use $C_0$ as the final value. The end-product of this process is a rearranged list of VMs.
\section{Preliminary Evaluation}
We evaluate \cmpi with a series of microbenchmarks from various communication backends and real-world applications that use \collectives. We represent speedup by comparing the performance we get from the best rank order and the worst rank order. We avoid comparing with the original rank order because it is random and unstable (Figure~\ref{fig:azringperformance}).

\subsection{Experimental Setup}
Our experiments are conducted on two public clouds, \azure and \ectwo. We enable network acceleration on both clouds and set TCP congestion control protocol to DCTCP. We include microbenchmarks that exercise ring, having doubling, double binary tree, and Bcube algorithms.  All experiments run on Ubuntu 19. We focus our evaluation on one of the most important tasks in \mpi, \textit{allreduce} for its popularity. %We use unmodified Facebook Gloo~\cite{glooalgo70:online}, OSU MPI microbenchmark~\cite{10.1007/978-3-642-33518-1_16} and Nvidia NCCL~\cite{NVIDIACo76:online} with OpenMPI 4. We use lightgbm~\cite{Ke2017LightGBMAH} to evaluate the real-world impact of \cmpi. 

\subsection{Prediction Accuracy of Cost Model}
While the goal of the cost model is not to predict the actual performance, but rather, it should preserve the relative order of performance, i.e., $p_{pred}(\mathbb{R}_1) < p_{pred}(\mathbb{R}_2) \implies p_{real}(\mathbb{R}_1) < p_{real}(\mathbb{R}_2)$ should hold true for as many pairs of $(R_i, R_j)$s as possible. We demonstrate this for ring based \mpi by generating 10 different rank orders, with the $i$-th order approximately corresponds to the $10i$-th percentile in the range of costs found by the solver. We obtain performance data for Facebook Gloo and OpenMPI running OSU Benchmark on 64 F16 nodes on \azure and 64 C5 nodes on \ectwo. We then compute Spearman~\cite{spearman} correlation coefficient between the predicted performance and the actual performance for each setup~(Table~\ref{table:correlation}). We found the cost function predicting and actual collectives performance exhibits strong correlation.

\begin{table}[t!]
    \centering
    \footnotesize
    \begin{tabular}{|c|c|c|}
        \hline 
        Setup & Azure & EC2  \\
        \hline
        Gloo Ring 100MB      & 0.58 & 0.78  \\
        \hline
        OpenMPI Ring 100MB      & 0.81 & 0.94 \\
        \hline
    \end{tabular}
    \caption{Spearman correlation coefficient between predicted performance from cost model and actual performance.}
    \label{table:correlation}
\end{table}

%Figure~\ref{fig:azringperformance} shows where the predicted best (purple dotted line) and worst performance (black dotted line) using the cost model landed in the performance distribution from 500 random ordering: the $\mathbb{R}$ that is predicted to achieve worst/best performance falls in the 99th/1st percentile of the total observed distribution. While not perfect, we believe this is acceptable due to the imperfection of the cost model and the limitation of solvers. 

\subsection{Microbenchmark Performance}
We evaluate \cmpi{}'s efficacy with microbenchmarks of algorithms introduced in~\ref{sec:background}. We report a mean speedup of 20 iterations. We run all benchmarks with 512 F16 nodes on \azure, except for NCCL, which runs on 64 P3.8xLarge GPU nodes on \ectwo. Specifically, we set $B=4$ for BCube; for NCCL, we use a single binary tree reduction for small buffers and a ring for large buffers. In all benchmarks, we reduce a buffer of 100MB, except for Nvidia NCCL, where we reduce a small buffer of 4B to trigger the tree algorithm.

\begin{figure}[h]
    \centering
    \includegraphics[width=\linewidth]{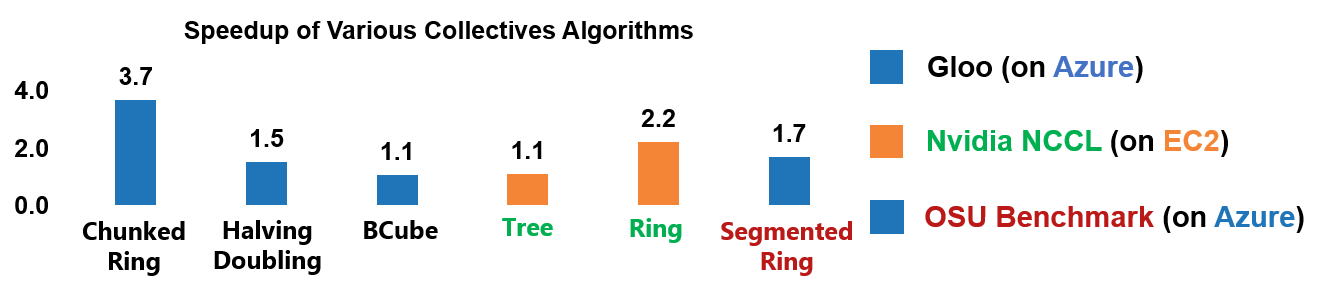}
    \caption{Speedups achieved by just using the rank ordering produced by our tool in various algorithms across multiple distributed communication backends at a large scale.}
    \label{fig:collectivesPerformance}
\end{figure}

Figure~\ref{fig:collectivesPerformance} shows a summary of speedups achieved using \cmpi on these benchmarks with the ring-family algorithms benefiting the most, up to 3.7x. We speculate the reason for the effectiveness is that they have a much wider performance distribution, as each permutation of the order can potentially generate a different performance (cost of each hop is on the critical path); they also have simpler cost model, allowing the solvers to quickly navigate the objective landscape. On the other hand, halving doubling, BCube, and tree algorithms have complex objectives -- sum of maximums, resulting in a narrower performance distribution because mutation of the ordering may not change the cost at all if the mutation does not cause the critical path to change. %This adds to difficulty of minimizing the objectives, because the solver itself relies on efficiently generating new cost values to proceed. 

\subsection{End-to-end Performance Impact on Real-world Applications}
\noindent \textbf{Distributed Gradient Boosted Decision Tree}. We evaluate \cmpi{}'s impact on LightGBM~\cite{Ke2017LightGBMAH}, a gradient boosted decision tree training system. We use data parallelism to run \textit{lambdarank} with metric \textit{ndcg}. Communication-wise, this workload runs two tasks: \textit{allreduce} and \textit{reducescatter}, and %Each call to \textit{allreduce} reduces 24B of data, and approximately 1KB for \textit{reducescatter}. 
they are called sequentially in each split of the iteration. At our scale of 512 nodes, LightGBM automatically chooses to use halving and doubling for both \textit{reducescatter} and \textit{allreduce}. We use a dataset that represents an actual workload in our commercial setting with 5K columns and a total size of 10GB for each node. We train 1K trees, each with 120 leaves. We exclude the time it takes to load data from disk to memory and report an average speedup of 1000 iterations. \cmpi generated rank orders speed up training by 1.3x.

\noindent \textbf{Distributed Deep Neural Network}. We show \cmpi{}'s effectiveness on distributed training of DNNs with Caffe2/Pytorch, on 64 EC2 p3.8xLarge nodes with data parallelism and a batch size of 64/GPU. We train AlexNet on the ImageNet dataset. Since our \cmpi{} does not change computation and only improves communication efficiency of the \textit{allreduce} operation at iteration boundary, we report speedup of training in terms of images/second, averaged across 50 iterations. We use the ring chunked algorithm, which achieves the best baseline performance, and the \cmpi{}-optimized rank ordering of VM nodes achieves a speedup of 1.2x.
\section{Discussion and Limitations}

\noindent\textbf{Generalizability Study.} Due to lack of resources, we only evaluated \cmpi on a few VM allocations on each cloud. Further investigation is needed to understand how \cmpi performs across different physical allocations and whether \cmpi{}'s rank order reacts to cloud variance well.

\noindent\textbf{Limitations to Cost Modeling.} Our use of latency-centric cost function is reasonable and effective, but is not perfect: it is possible that different bandwidth can correspond to the same latency depending on link condition. This may cause unoptimal solution when a transfer is bandwidth-bound. Further study is needed to determine how to incorporate bandwidth related cost into the cost function without incurring high probing overhead.

\noindent\textbf{Complements to Cost Models}. While building accurate cost models is difficult, we can dynamically adjust rank ordering with help from tools such as TCP\_INFO~\cite{mathis2003web100} that monitors link properties such as latency and bandwidth. Since we know the communication pattern, we can determine the critical path and find bottleneck transfer between node $n_i$ and $n_j$ in the system. From there, we can find a $n_k$ to replace $n_i$ such that the replacement results in a minimized cost objective.

\noindent\textbf{Adapting to Dynamic Traffic}.
The above mechanism can be applied to adapt to dynamic network load in the cloud environment. The framework, however, must support the dynamic change of node ranks, which is possible and should come with a small cost as a full mesh of connections can be established beforehand among all nodes.

%%%%%%% -- PAPER CONTENT ENDS -- %%%%%%%%

%%%%%%%%% -- BIB STYLE AND FILE -- %%%%%%%%
\bibliographystyle{IEEEtranS}
\bibliography{paper}
%%%%%%%%%%%%%%%%%%%%%%%%%%%%%%%%%%%%

\end{document}